\documentclass{jpconf}

\usepackage{amsmath,amssymb}

\newcommand{\beq}{\begin{equation}}
\newcommand{\eeq}{\end{equation}}
\newcommand{\beqa}{\begin{eqnarray}}
\newcommand{\eeqa}{\end{eqnarray}}
\newcommand{\nn}{\nonumber}
\newcommand{\f}{\frac}

\newcommand{\mf}{\mathfrak}
\newcommand{\mbb}{\mathbb}
\newcommand{\mc}{\mathcal}

\newcommand{\dzeta}{\zeta_{\mathbb{Q}(i\sqrt{d})}}

\begin{document}

\title{Rigid Calabi-Yau threefolds, Picard Eisenstein series and instantons}

\author{L Bao${}^1$, A Kleinschmidt${}^2$,
B E W Nilsson${}^3$, D Persson${}^4$ and B Pioline${}^5$}
\address{${}^1$ Institut des Hautes Etudes Scientifiques, Le Bois-Marie 35, 91440 Bures-sur-Yvette,
France}
\address{${}^2$ Physique Th\'eorique et Math\'ematique, Universit\'e Libre de Bruxelles and International Solvay Institutes, Campus Plaine CP 231, Boulevard du Triomphe, 1050 Brussels, Belgium}
\address{${}^3$ Fundamental Physics, Chalmers University of Technology, 412 96, G\"oteborg, Sweden}
\address{${}^4$ Institut f\"ur Theoretische Physik, Eidgen\"ossische Technische Hochschule, 8093 Z\"urich, Switzerland}
\address{${}^5$ Laboratoire de Physique Th\'eorique et Hautes Energies, CNRS UMR 7589 and Universit\'e Pierre et Marie Curie - Paris 6, 4 place Jussieu, 75252 Paris cedex 05, France}
\ead{lingbao@ihes.fr,axel.kleinschmidt@ulb.ac.be,tfebn@chalmers.se,\\daniel.persson@itp.phys.ethz.ch,pioline@lpthe.jussieu.fr}
\begin{abstract}
Type IIA string theory compactified on a rigid Calabi-Yau threefold gives rise to a classical moduli space that carries an isometric action of $U(2,1)$. Various quantum corrections break this continuous isometry to a discrete subgroup. Focussing on the case where
the intermediate Jacobian of the Calabi-Yau admits complex multiplication by the ring of quadratic
imaginary integers $\mc{O}_d$, we argue that the remaining quantum duality group is an arithmetic Picard modular group $PU(2,1;\mc{O}_d)$. Based on this proposal we construct an Eisenstein series invariant under this duality group and study its non-Abelian Fourier expansion. This allows the prediction of non-perturbative effects, notably the contribution of D2- and NS5-brane instantons. The present work extends our previous analysis in 0909.4299 which was restricted to the special case of the Gaussian integers $\mathcal{O}_1=\mathbb{Z}[i]$.
\end{abstract}

\section{Introduction}

In this contribution we are interested in constructing the exact hypermultiplet moduli space metric of type IIA string theory compactified on a rigid Calabi-Yau threefold $\mc{X}$, admitting complex multiplication by the ring of quadratic imaginary integers $\mathcal{O}_d$. The rigidity condition means that there are no complex structure deformations ($h_{2,1}(\mc{X})=0$) and therefore the hypermultiplet sector of the resulting $\mc{N}=2$ theory in $D=3+1$ dimensions consists only of the so-called universal hypermultiplet, parameterized by four scalar fields.  The complex multiplication property implies that the intermediate Jacobian of $\mc{X}$ takes the form
\beq
\mc{J}_d=H^{3}(\mc{X},\mbb{C})/H^{3}(\mc{X},\mbb{Z})\simeq \mbb{C}/\mc{O}_d\,,
\eeq
where $\mathcal{O}_d$ is the set of algebraic integers in the quadratic number field
$\mbb{Q}(\sqrt{-d})$ where $d$ is a square-free positive integer. For simplicity we restrict to the case where $\mathcal{O}_d$ is a principal ideal domain, hence admitting unique prime factorization. Remarkably, according to the Stark-Heegner theorem, there are only a finite number of values of $d$‚ which satisfy this criterion:
\beq\label{shseq}
d\in \{ 1,2,3,7,11,19,43,67,163\}\,.
\eeq
The case $d=1$ corresponds to the Gaussian integers $\mc{O}_1=\mathbb{Z}+i\mathbb{Z}$ and was treated in detail in~\cite{Bao:2009fg}, where more details on the physical motivations behind the set-up studied in this contribution can be found. Here, we restrict exclusively to the case $d\equiv 3 \mod 4$ -- which covers the entire Stark-Heegner sequence except $d=1,2$ -- in which case
\beq
\mc{O}_d\equiv \mbb{Z}[\omega_d]=\mathbb{Z}+\omega_d\mathbb{Z}, \quad\quad\text{where}\quad \quad  \omega_d = \frac{-1+i\sqrt{d}}{2}\,.
\eeq

We parameterize the hypermultiplet moduli space $\mc{M}_{\text{UH}}$ by four scalars, namely the 4D dilaton $\phi$ and the axions $\zeta,\tilde\zeta,\sigma$. The scalars $\zeta$ and $\tilde\zeta$ are the periods of the Ramond-Ramond 3-form along a symplectic basis of $H_3(\mathcal{X},\mathbb{Z})$, and the complex variable $\zeta+\omega_d \tilde\zeta$ thus parameterizes the intermediate Jacobian $\mathcal{J}_d$. The coordinate $\sigma$, on the other hand, corresponds to the $D=4$ dual of the Neveu-Schwarz two-form, and parameterizes the fiber of a circle bundle over $\mathcal{J}_d$. The classical metric on $\mc{M}_{\text{UH}}$ can be written as
\beq
ds_{\mc{M}_{\text{UH}}}^{2}=d\phi^2+\f{1}{2} e^{2\phi}
\frac{|d\zeta+\omega_d \, d\tilde\zeta|^2}{\Im({\omega_d})}+\f{1}{4}e^{4\phi}\Big(d\sigma-\tilde\zeta d\zeta+\zeta d\tilde\zeta\Big)^2\,,
\label{UHMmetric}
\eeq
which can be derived via the $c$-map from the quadratic prepotential $F(X)=\omega_d X^2/2$ on the type IIB side, with $\omega_d$ playing the role of the ``period matrix''. The metric (\ref{UHMmetric}) is locally isometric to the quaternion-K\"ahler symmetric space $(U(2)\times U(1))\backslash U(2,1)$.
Quantum corrections break the continuous isometry group $U(2,1)$ to a discrete
subgroup $\Gamma$, while preserving the quaternion-K\"ahler  property. $\Gamma$ should contain at least the Heisenberg group
\beq\label{shiftgp}
\zeta  \longrightarrow  \zeta +a\,,\quad
\tilde\zeta  \longrightarrow  \tilde\zeta +b\,,\quad
\sigma  \longrightarrow  \sigma +2c-a \tilde\zeta+b\zeta +(a+b+ab)\,,
\eeq
for $a,b,c\in\mbb{Z}$. These discrete periodicities are remnants of  continuous (Peccei-Quinn) symmetries of the metric (\ref{UHMmetric}), as a result of D2 and
NS5-brane instanton effects~\cite{Becker:1995kb,Becker:1999pb}. In particular, the last term in brackets in the shift of $\sigma$ is consistent with the quadratic refinement appearing in the NS5-brane partition function~\cite{Belov:2006jd,APP}.
Examining the list of generators of the Picard modular groups~\cite{FalbelParker,GaussianPicard,Jiang,Bao:2009fg},  it is natural to
conjecture that
$\Gamma$ is the {\em Picard modular group} $PU(2,1;\mc{O}_d)\equiv U(2,1)\cap PGL(3,\mc{O}_d)$, in the standard embedding. It contains (\ref{shiftgp}) as a unipotent subgroup $N(\mc{O}_d)=N\cap \Gamma$, where $N$ is associated to the Iwasawa decomposition $U(2,1)=KAN$. The additional generators in  $\Gamma$ can be identified physically as electric-magnetic duality and S-duality. Whereas the discrete symmetries (\ref{shiftgp}) are well-established, the presence of electric-magnetic duality and S-duality at the quantum level is a strong assumption which merits further justification.\footnote{It is debatable whether the correct quantum duality group ought to be $PU(2,1;\mc{O}_d)$ or $PSU(2,1;\mc{O}_d)$. Except for the $d=1$ case discussed in~\cite{Bao:2009fg}, the latter group, unlike the former, does not include electric-magnetic duality.  The simplest Eisenstein series constructed below is in fact invariant under the full $PU(2,1;\mc{O}_d)$.}  In~\cite{Bao:2009fg} we have given some arguments in favour of this assertion and refer the reader to the discussion there.

Our aim in the remainder will be to construct a $\Gamma$-invariant automorphic form on the coset space $\mc{M}_{\text{UH}}$, namely the Eisenstein series $\mc{E}_s$ attached to a  degenerate principal continuous representation of $U(2,1)$. The ultimate goal is to extract the {\em exact} metric on $\mc{M}_{\text{UH}}$ by using twistor techniques for quaternion-K\"ahler manifolds. The basic idea is that deformations of any quaternion-K\"ahler manifold $\mathcal{M}$ can be encoded in deformations of the complex contact structure of the associated twistor space $\mathcal{Z}_\mc{M}$, which is a $\mathbb{C}P^1$-bundle over $\mathcal{M}$ carrying a K\"ahler-Einstein metric \cite{Salamon,LeBrun}. The K\"ahler potential $K_{\mathcal{Z}_{\mathcal{M}}}$ for the K\"ahler metric on $\mathcal{Z}_{\mathcal{M}}$ follows from the so called {\em contact potential} $\Phi(x,z)$, which depends on the collective coordinates $x$ on the base $\mathcal{M}_{\text{UH}}$ and (holomorphically) on the fiber-coordinate $z\in \mathbb{C}P^1$, through the formula $K_{\mathcal{Z}}=\log \big[(1+|z|^2)/|z|\big]+ \Re(\Phi)$. The knowledge of the contact potential together with the so-called twistor lines -- relating local Darboux coordinates on $\mc{Z}_\mc{M}$ to the pair $(x,z)\in \mathcal{M}\times \mathbb{C}P^1$ --  then completely determines the metric on $\mc{M}_{\text{UH}}$ \cite{Alexandrov:2008nk,Alexandrov:2008gh}.  Our tentative proposal is that the exact contact potential at the `north pole' $z=0$ of the twistor space is given by
\beq\label{propcont}
e^{\Phi} = e^\phi\,  \mc{E}_{s=3/2}\,,
\eeq
where $\phi$ on the r.h.s.~is the dilaton and the order $s=3/2$ of the Eisenstein series is fixed by studying the known perturbative terms. To completely specify the metric on
$\mc{M}_{\text{UH}}$, one should also determine the quantum corrections to the twistor lines,
which were derived at the classical level  in~\cite{Gunaydin:2007qq}.
To motivate our proposal further we will now construct the Eisenstein series $\mc{E}_s$ and analyze its Fourier expansion with respect to the Heisenberg group (\ref{shiftgp}).

\section{Construction of Eisenstein series for $PU(2,1;\mc{O}_d)$}

It will be useful to exploit the isomorphism between the coset space $(U(2)\times U(1))\backslash U(2,1)$ and complex hyperbolic space $\mc{U}$, where the latter is defined in terms of a pair of complex coordinates $\mc{Z}=(z_1,z_2)$ as follows
\beq
\mc{U}= \left\{ \mc{Z}=(z_1,z_2)\in\mbb{C}^2\,:\,\mc{F}(\mc{Z})= -\Re(z_1) - \frac12 |z_2|^2 >0\right\}\,.
\eeq
This space is preserved by the standard fractional linear action of the continuous $U(2,1)$, defined as the set of matrices  leaving invariant the indefinite metric
\beq
\eta=\left(\begin{array}{ccc}
0& 0& 1\\
0& 1 & 0\\
1 & 0&0\\
\end{array}\right)\,,\quad\quad U(2,1)=\left\{g \in GL(3,\mbb{C})\,:\, g^\dagger\eta g=\eta\right\}\,.
\eeq
The K\"ahler metric on $\mc{U}$ can be derived from the K\"ahler potential $K=-\log(\mc{F})$, leading to
\beq\label{Kmetric}
ds^2 = \frac14 \mc{F}^{-2} \big( dz_1 d\bar{z}_1 +z_2 dz_1 d\bar{z}_2 + \bar{z}_2 dz_2 d\bar{z}_1-2\Re(z_1) dz_2 d\bar{z}_2\big) \,.
\eeq
This becomes identical to the metric (\ref{UHMmetric}) upon identifying
\beq
z_1 \equiv i \Im(\omega_d)\sigma
-\f{1}{2}|\zeta+\omega_d\tilde\zeta|^2-\Im(\omega_d) e^{-2\phi}\,,\quad
z_2\equiv \zeta+\omega_d \tilde\zeta\,,
\label{changevar}
\eeq
so that $\mc{F}=\Im(\omega_d)\, e^{-2\phi}$.  In the following it will prove convenient to work directly with the complex
 coordinates $(z_1,z_2)$ rather than the real coordinates $(\phi, \zeta, \tilde\zeta, \sigma)$.

The metric (\ref{Kmetric}) can alternatively be written as $ds^2=-\frac18 \text{Tr}(d\mc{K} d\mc{K}^{-1})$ in terms of the  $U(2)\times U(1)$-invariant Hermitean coset representative
\beq
\mc{K} = \mc{F}^{-1}\left(\begin{array}{ccc}
1 &  z_2  & z_1 \\
 \bar{z}_2 & |z_2|^2 &\bar{z}_2 z_1 \\
\bar{z}_1 &\bar{z}_1 z_2 & |z_1|^2\end{array}\right)
+\left(\begin{array}{ccc}
0 & 0& 1 \\
0& 1 & 0 \\
1 &0 &0\end{array}\right)\,\in SU(2,1)\ .
\eeq
Defining also the singular matrix $\tilde{\mc{K}}=\mc{K}-\eta$, one sees that $\tilde{\mc{K}}$ can be written as
\beq\label{tKfact}
\tilde{\mc{K}}= \tilde{\mc{V}}^\dagger \tilde{\mc{V}}\quad\quad\text{with}\quad
\tilde{\mc{V}} = \mc{F}^{-1/2} \left(\begin{array}{ccc}
1 & z_2 & z_1\\
0 & 0 & 0\\
0 & 0 & 0\end{array}\right)\,.
\eeq

A $\Gamma$-invariant Eisenstein series $\mc{E}_s(\mc{Z})$ on $\mc{U}$ can then be defined as a (constrained) sum over non-zero lattice vectors $\vec\Omega=(\Omega_1,\Omega_2,\Omega_3)^T\in\mc{O}_d^3$ according to
\beq\label{eisendef}
\mc{E}_s (\mc{Z}) = \sum_{0\neq \vec\Omega\in \mc{O}_d^3\atop \vec\Omega^\dagger\eta\vec{\Omega}=0} (\vec\Omega^\dagger \mc{K} \vec\Omega)^{-s}
= \mc{F}^s  \sum_{0\neq \vec\Omega\in \mc{O}_d^3\atop \vec\Omega^\dagger\eta\vec{\Omega}=0}
|\Omega_3 z_1 +\Omega_2 z_2 + \Omega_1|^{-2s}\,.
\eeq
The constraint
\beq
\label{cons}
\vec\Omega^{\dagger}\, \eta\, \vec\Omega=|\Omega_2|^2+2\Re(\Omega_1\bar\Omega_3) =0
\eeq
on the lattice vectors is necessary to ensure that $\mc{E}_s$ is an eigenfunction of the $PU(2,1)$ invariant Laplacian on $\mc{U}$. For vectors $\vec\Omega$ satisfying the constraint (\ref{cons}) one can replace $\mc{K}$ by $\tilde{\mc{K}}$ and then use the factorization (\ref{tKfact}) to rewrite the sum as shown in the last equality of (\ref{eisendef}).

When performing the Fourier expansion of $\mc{E}_s$ it is expedient to solve the constraint on $\vec{\Omega}$ explicitly. To this end we note that for fixed  $\Omega_2, \Omega_3\in \mathcal{O}_d$ with $\Omega_3\neq 0$, solutions to (\ref{cons}) exist if and only if $|\Omega_2|^2/r\in\mbb{Z}$, where $r$ is defined as
\beq\label{rdef}
r= \gcd\left(2p_1-p_2, \frac{d+1}{2}p_2 -p_1\right)\ ,\qquad
\Omega_3\equiv p_1+\omega_d p_2\ ,\qquad p_1,p_2\in\mathbb{Z} \,.
\eeq
In this case, the solutions of (\ref{cons}) can be parameterized as
\beq\label{conssol}
\Omega_1 = - \frac{|\Omega_2|^2}{r} \Omega_1^{(0)} - i\sqrt{d} \frac{m}{r} \Omega_3\quad\quad m\in\mbb{Z} \,,
\eeq
where $\Omega_1^{(0)}\in \mathcal{O}_d$ is a particular solution of $r=2\Re(\Omega_1^{(0)}\bar{\Omega}_3)$, obtained via the Euclidean algorithm.

\section{Abelian and non-Abelian Fourier coefficients}

Since the unipotent subgroup $N(\mc{O}_d)\subset \Gamma$ is a three-dimensional Heisenberg group, the general Fourier expansion of a $\Gamma$-invariant Eisenstein series consists of three distinct pieces: the so-called constant, Abelian and non-Abelian terms~\cite{Ishikawa,Pioline:2009qt}. The constant terms correspond to the zeroth Fourier coefficients, and hence do not depend on the axions; these terms can therefore be identified with perturbative contributions in the string coupling $g_s=e^{\phi}$. Due to the fact that the (restricted) Weyl group of $U(2,1)$ is of order 2, one expects two constant terms~\cite{Langlands}. The axion-dependent terms (generic Fourier coefficients) further separate into Abelian and non-Abelian parts, respectively corresponding to the expansion along $N/Z$ and $Z$, where $Z$ is the center of $N$. We will now extract these pieces in turn by performing appropriate Poisson resummations.

To begin with, the Eisenstein series (\ref{eisendef}) can be separated into the terms where $\Omega_3=0$ (implying $\Omega_2=0$ by (\ref{cons})) and those where $\Omega_3\neq 0$ as
\beq
\mc{E}_s(\mc{Z}) = \mc{E}_s^{(0)}(\mc{F}) + \mc{A}_s(\mc{Z})\,.
\eeq
The term $\mc{E}_s^{(0)}$ is the leading order contribution in the weak coupling limit $e^{\phi}\rightarrow 0$ and  evaluates straightforwardly as
\beq
\label{cons1}
\mc{E}_s^{(0)} = \mc{F}^s \sum_{\Omega_1\neq0} |\Omega_1|^{-2s} = \mc{F}^s \dzeta(s)\,,
\eeq
where we have introduced the Dedekind zeta function $\dzeta(s)$. Using the structure of primes in the ring $\mc{O}_d$, it can be shown that
\beq
\dzeta(s)=e_d\, \beta_d(s)\, \zeta(s)
\eeq
where $e_d$ is the number of units in $\mc{O}_d$  (i.e. $e_3=6$ and $e_d=2$ otherwise),
$\beta_d(s)\equiv L\left( {\tiny \left( \begin{array}{c} \cdot \\ \hline d \end{array}\right)},s \right)$
is the Dirichlet L-function associated to the Legendre symbol and $\zeta(s)$ is the ordinary Riemann zeta function.
$\mc{E}_s^{(0)}$ represents the first of the two constant terms.

Turning to the remainder term $\mc{A}_s$, we can now use the explicit solution of the constraint given in (\ref{conssol}). Furthermore, we employ some standard tricks for the rewriting of powers as
\beq
M^{-s} = \frac{\pi^s}{\Gamma(s)} \int_0^\infty \frac{dt}{t^{s+1}} e^{-\frac{\pi}{t} M}\quad\quad\text{for $M>0$}
\label{IntegralExpr}
\eeq
and for Poisson resummation (for real $x>0$ and complex $a=a_1+i a_2$)
\beq
\sum_{m\in\mbb{Z}} e^{-\pi x |m+a|^2} = \frac{1}{\sqrt{x}}\sum_{\tilde{m}\in\mbb{Z}} e^{-\frac{\pi}{x} \tilde{m}^2 -2\pi i \tilde{m}a_1 -\pi x a_2^2}\,.
\eeq
Using this leads to
\beqa\label{fulla1}
\mc{A}_s &=& \mc{F}^s \sum_{\Omega_3\neq 0} \sum_{\Omega_2\in \mc{O}_d\atop
|\Omega_2|^2 \in r\mbb{Z}} \sum_{m\in\mbb{Z}} \left|\Omega_3z_1 +\Omega_2 z_2 -\frac{|\Omega_2|^2}{r}\Omega_1^{(0)}-i\sqrt{d}\frac{m}{r}\Omega_3\right|^{-2s}\nn\\
&=& \mc{F}^{s} \frac{\pi^s}{\Gamma(s)} \sum_{\Omega_3\neq0} \sum_{\Omega_2\in \mc{O}_d\atop|\Omega_2|^2\in r \mbb{Z}}\frac{r}{|\Omega_3|\sqrt{d}} \sum_{\tilde{m}\in\mbb{Z}}
\int_0^\infty \frac{dt}{t^{s+1/2}} e^{-\pi t \frac{r^2}{d|\Omega_3|^2} \tilde{m}^2 - 2\pi i \tilde{m} a_1 - \pi \frac{|\Omega_3|^2 d}{r^2 t}a_2^2}\,,
\eeqa
where
\beqa
a_1 &=& -\frac{r}{\sqrt{d}|\Omega_3|^2}\left(|\Omega_3|^2\Im(z_1) +\Im(\Omega_2\bar{\Omega}_3z_2)-\frac{|\Omega_2|^2}{r}\Im(\bar{\Omega}_3\Omega_1^{(0)})\right)\,,\nn\\
a_2 &=& -\frac{r}{\sqrt{d}|\Omega_3|^2} \bigg(|\Omega_3|^2 \mc{F} + \frac12|\Omega_2-\Omega_3\bar{z}_2|^2\bigg)\,.
\eeqa

The second constant term and the Abelian Fourier coefficients can be extracted from the $\tilde{m}=0$ terms, leading to
\beqa
\mc{A}_s^{(\tilde{m}=0)} &=& \mc{F}^{s} \frac{\pi^s}{\Gamma(s)} \sum_{\Omega_3\neq0} \sum_{\Omega_2\in \mc{O}_d\atop
|\Omega_2|^2\in r \mbb{Z}}\frac{r}{|\Omega_3|\sqrt{d}}
\int_0^\infty \frac{dt}{t^{s+1/2}} e^{ - \pi \frac{|\Omega_3|^2 d}{r^2 t}a_2^2}\\
&=& \mc{F}^{s} \frac{\pi^s\Gamma(s-1/2)\pi^{2s-1}}{\Gamma(s)\pi^{s-1/2}\Gamma(2s-1)} \sum_{\Omega_3\neq0} \sum_{\Omega_2\in \mc{O}_d\atop|\Omega_2|^2\in r\mbb{Z}}\frac{r}{\sqrt{d}|\Omega_3|^{2-2s}}
\int_0^\infty \frac{dt}{t^{2s}} e^{-\frac{\pi}{t}\left(|\Omega_3|^2\mc{F}+\frac12 \left|\Omega_2-\Omega_3\bar{z}_2\right|^2\right)}\,, \nn
\eeqa
where, in getting from the first to the second line, we have performed the integral over $t$  and then used (\ref{IntegralExpr}) to re-exponentiate part of the summand.

In the next step we have to treat the divisibility constraint $|\Omega_2|^2 \in r \mathbb{Z}$. This can be resolved by writing the solution as
\beq
\Omega_2 = \Omega_2^{(0)} + r v\,,
\eeq
where $v\in\mc{O}_d$, and $\Omega_2^{(0)}$ runs over the set $F_r$ of solutions to the equation $|\Omega_2^{(0)}|^2 = 0 \ \text{mod}\ r$ in a fundamental domain $\left\{n_1+n_2 \omega_d\;:\; 0\leq n_1,n_2 <r\right\}$, with cardinality $N_d(r) \equiv \# F_r$. Now one can
Poisson resum over $v\in\mc{O}_d$ using
\beq
\sum_{v\in \mc{O}_d} e^{-\pi x |v+a|^2} = \frac2{x\sqrt{d}} \sum_{u\in\mc{O}_d^*} e^{-\frac{\pi}{x}|u|^2-2\pi i \Re(u\bar{a})}\,,
\eeq
where $u$ runs over the dual lattice $\mc{O}_d^*$, spanned by $u_1\equiv \tfrac{2}{d}\left(\tfrac{d+1}{2}+\omega_d\right)=1+\tfrac{i}{\sqrt{d}}$ and $u_2\equiv \tfrac{2}{d}\left(1+2\omega_d\right)=\tfrac{2i}{\sqrt{d}}$.
The resulting Abelian terms are
\beqa\label{ab1}
\mc{A}_s^{(\tilde{m}=0)}&=& \mc{F}^{s} \frac{4}{d}\frac{\pi^s\Gamma(s-1/2)\pi^{2s-1}}{\Gamma(s)\pi^{s-1/2}\Gamma(2s-1)} \sum_{\Omega_3\neq0} \sum_{\Omega_2^{(0)}\in F_r}\frac{1}{r |\Omega_3|^{2-2s}} \nn\\
&&\times \sum_{u\in\mc{O}_d^*} \int_{0}^\infty \frac{dt}{t^{2s-1}} e^{-\frac{\pi}{t}|\Omega_3|^2\mc{F}-\frac{2\pi t}{r^2}|u|^2 -2\pi i\frac1r \Re(u(\bar{\Omega}_2^{(0)}-\bar{\Omega}_3z_2))}.
\eeqa
The second constant term arises from  the $u=0$ part of the sum:
\beq
\label{cons2}
\mc{A}_s^{(\tilde{m}=u=0)} = \mc{F}^{2-s} \frac{4}{d}\frac{\pi^s\Gamma(s-1/2)\pi^{2s-1}\Gamma(2s-2)}{\Gamma(s)\pi^{s-1/2}\Gamma(2s-1)\pi^{2s-2}} \sum_{\Omega_3\neq0} \sum_{\Omega_2^{(0)}\in F_r}\frac{1}{r |\Omega_3|^{2s-2}}.
\eeq

Noting that $r$ divides $|\Omega_3|^2$, and using similar techniques as in~\cite{Bao:2009fg}
to sum up the  Dirichlet series $\sum_{r>0} N_d(r) r^s$, one can rewrite the sums
over $\Omega_3$
and $\Omega_2^{(0)}$ as follows:
\beq
 \sum_{\Omega_3\neq0} \sum_{\Omega_2^{(0)}\in F_r}\frac{1}{r |\Omega_3|^{2s-2}}=\left( \sum_{r>0} N_d(r) r^{1-2s} \right) \frac{\dzeta(s-1)}{\zeta(2s-2)}=\frac{\beta_d(2s-2)}{\beta_d(2s-1)}\dzeta(s-1),
 \eeq

Orchestrating, the second constant term (\ref{cons2}) becomes
\beq\label{cons3}
\mc{A}_s^{(\tilde{m}=u=0)} = \mc{F}^{2-s} \dzeta(s) \frac{\mf{Z}_d(2-s)}{\mf{Z}_d(s)}\,,
\eeq
where we have defined the `completed Picard function'
\beq
\mf{Z}_d(s) = \dzeta^*(s) \beta_d^*(2s-1)
\eeq
in terms of the completed Dedekind zeta and Dirichlet L-functions,
\beqa
\dzeta^*(s) &=&\left(\frac{d}{4} \right)^{s/2} \pi^{-s}\, \Gamma(s) \, \dzeta(s)\,,\nn\\
\beta_d^*(s) &=& \left(\frac{d}{4}\right)^{(s+1)/2}\left(\frac{\pi}{4}\right)^{-(s+1)/2}\Gamma\left(\frac{s+1}{2}\right)\,\beta_d(s)\, ,
\eeqa
both being invariant under the interchange $s\leftrightarrow 1-s$. The two constant terms
(\ref{cons1}) and (\ref{cons3}) can be now summarized as
\beqa\label{constboth}
\mc{E}_s^{(\text{const})}(\mc{Z}) = \dzeta(s)\left[ \mc{F}^s +  \frac{\mf{Z}_d(2-s)}{\mf{Z}_d(s)} \mc{F}^{2-s}\right]\,.
\eeqa

The Abelian terms correspond to the terms in (\ref{ab1}) with $u\neq 0$. The integral over
$t$ produces a sum of modified Bessel functions,
\beqa\label{fourab}
\mc{A}_s^{(\tilde{m}=0,\,u\neq0)} &=& \mc{F} \frac{2^{s+2}}{d}\frac{\pi^s\Gamma(s-1/2)\pi^{2s-1}}{\Gamma(s)\pi^{s-1/2}\Gamma(2s-1)}\sum_{\Omega_3\neq0} \sum_{\Omega_2^{(0)}\in F_r}r^{1-2s} \sum_{u\in\mc{O}_d^*} |u|^{2s-2} e^{-2\pi i \frac1r \Re(u\bar{\Omega}_2^{(0)})}\nn\\
&&\times e^{2\pi i\frac1r \Re(u\bar{\Omega}_3z_2)} K_{2s-2}\left(2\pi \mc{F}^{1/2} \left|\frac{\sqrt{2}\bar{u}\Omega_3}{r}\right|\right).
\eeqa
Holding $\Lambda\equiv \bar{u}\Omega_3/r$ fixed
and carrying out the sum over $u\in \mc{O}_d^*$, $\Omega_3\in\mc{O}_d$
and $\Omega_2^{(0)}\in F_r$, this may be rewritten using the variable change
(\ref{changevar}) as
\beq
\mc{A}_s^{(\tilde{m}=0,\,u\neq0)} = e^{-2\phi}\frac{\dzeta(s)}{\mf{Z}_d(s)}
\sum_{\Lambda\neq 0} \mu_s(\Lambda) \, e^{2\pi i (p\zeta-q \tilde\zeta)}
K_{2s-2}\left(2\pi \sqrt2\, e^{-\phi}\frac{|q+\omega_d \, p|}{\sqrt{\Im(\omega_d)}}\right)\,,
\label{AbelianTerm}
\eeq
for a summation measure $\mu_s(\Lambda)$ with $\Lambda$ expanded in the basis $(u_1,u_2)$ of $\mc{O}_d^*$, namely
\beq
\Lambda= \frac{\bar{u}\Omega_3}{r} \equiv p u_1-q u_2 =
-i \frac{(q + \omega_d\, p)}{\Im(\omega_d)}\ .
\eeq
The fact that $\mu_s(\Lambda)\equiv\mu_s(p,q)$ has support on integer charges $p,q\in\mathbb{Z}$ is not obvious from their definition, but is required by the periodicity conditions (\ref{shiftgp}) and results from
cancellations in the sum over $\Omega_2^{(0)}$. Hence only those $\Omega_3$ for which $r=\gcd(p_1,p_2)$ contribute and we write $\Omega_3'=\Omega_3/\gcd(p_1,p_2)$ for the primitive vector in $\mc{O}_d$. Using similar methods  that were
employed in~\cite{Bao:2009fg} for the Gaussian integers $\mathcal{O}_1=\mathbb{Z}[i]$, it is possible to express  $\mu_s(\Lambda)$ as a double divisor sum if $d$ belongs to the Stark--Heegner sequence (\ref{shseq}) (up to an $s$-independent factor) as
 \beq
 \mu_s(\Lambda) = \sum_{\Omega_3'\in \mc{O}_d\atop {\Lambda}/{\Omega_3'}\in\mc{O}_d^*} \left|\frac{\Lambda}{\Omega_3'}\right|^{2s-2} \sum_{z\in\mc{O}_d\atop{\Lambda}/{z\Omega_3'}\in\mc{O}_d^*} |z|^{4-4s}\,.
 \eeq

Finally, the non-Abelian terms correspond to the contributions in (\ref{fulla1}) with $\tilde{m}\neq0$. These can also be simplified along the lines of~\cite{Bao:2009fg} but we refrain from displaying the end result here. In view of the constant terms (\ref{constboth}), it is natural to conjecture that  the Poincar\'e series $\mc{P}_s(\mc{Z}) \equiv \mc{E}_s(\mc{Z})/ \dzeta(s)$ for any $d$ in the Stark-Heegner sequence (\ref{shseq}) satisfies the functional relation (generalizing a statement put forward in~\cite{Bao:2009fg})
\beq
\mf{Z}_d(s)\, \mc{P}_s(\mc{Z}) = \mf{Z}_d(2-s)\, \mc{P}_{2-s}(\mc{Z})\,.
\eeq

\section{Discussion}

We now turn to a discussion of the various terms. The constant terms, summarized in (\ref{constboth}), are interpreted as corrections to the classical hypermultiplet  metric (\ref{UHMmetric}) that arise from string perturbation theory. Two different powers of the string coupling $g_s=e^{\phi}=\mc{F}^{-1/2} \sqrt{\Im(\omega_d)}$ occur in this expansion. Referring back to our proposal for the contact potential (\ref{propcont}) we see that  for the value $s=3/2$ of the order of the Eisenstein series $\mc{E}_s$, these two terms can be matched with string tree level $g_s^{-2}=e^{-2\phi}$ and one loop $g_s^0$ contributions. The relative coefficient between these two terms is known from explicit string theory computations (see for example~\cite{Antoniadis:2003sw}) and unfortunately  does not match the numerical value predicted by the ratio of completed Picard functions in (\ref{constboth}). 
Thus, our proposal does not work as such. Despite this, we shall nevertheless see that the remaining terms in the Fourier expansion match remarkably well with the expected form of instanton contributions to the hypermultiplet metric.

Having fixed the value of $s=3/2$ from the perturbative terms we can extract the weak-coupling ($e^\phi\rightarrow 0$ or $\mc{F}\rightarrow \infty$) limit from the Abelian terms. Examining (\ref{fourab}) one deduces that these terms are weighted by $e^{- S_{\text{D}2}}$, where ($u\neq 0$)
\beq
S_{\text{D}2} = 2\pi \sqrt{2}\, e^{-\phi} \frac{|q+\omega_d p|}{\sqrt{\Im(\omega_d)}} -
2\pi i (p\zeta-q \tilde\zeta) \,.
\eeq
This is recognized as the instanton action for Euclidean D2-branes wrapped on an arbitrary 3-cycle $\Lambda\in H_3(\mathcal{X},\mathbb{Z})$. The real part contains the expected $1/g_s$-suppression and the imaginary part exhibits the correct axionic ``theta angles''. The instanton measure $\mu_{3/2}(\Lambda)$ in (\ref{AbelianTerm}) should count special Lagrangian 3-cycles in the rigid Calabi-Yau threefold $\mc{X}$.

A similar analysis can be performed for the non-Abelian terms. This we do for simplicity for vanishing $\Omega_2=0$ which corresponds to pure NS5-brane instantons.  One may then infer from (\ref{fulla1}) that the leading contributions are weighted by $e^{-S_{\text{NS}5} }$, with (for $\tilde{m}\neq 0$)
\beq
S_{\text{NS}5} = \pi\, |r\tilde{m}|\, \left( e^{-2\phi} +\frac12
\frac{|\zeta+\omega_d \,\tilde\zeta|^2}{\Im(\omega_d)}
 \right)
-\pi i r\tilde{m}\sigma\, ,
\eeq
which may be identified with the action of a charge $r\tilde{m}$ Euclidean NS5-brane \cite{Becker:1995kb}. We note in particular that the real part is suppressed by $1/g_s^2$ which is a characteristic feature of NS5-brane instantons.  See \cite{Bao:2009fg} for a more detailed discussion when $d=1$.

Although our qualitative results are on the right track, as mentioned above the numerical predictions are incorrect. This indicates that our assumptions may be reasonable but that the construction requires some modification. Drawing inspiration from related work on implementing $SL(2,\mathbb{Z})$-invariance on the type IIB hypermultiplet moduli space~\cite{Alexandrov:2009qq}, it is natural to expect that the correct $PU(2,1;\mathcal{O}_d)$-invariant automorphic form should be a holomorphic function on the twistor space $\mathcal{Z}_\mc{M}$. Mathematically, 
this implies that the relevant automorphic form should be attached to the quaternionic discrete series of $U(2,1)$ (see \cite{GrossWallach,Gunaydin:2007qq}), rather than the principal continuous series considered here. Steps towards a twistorial formulation of NS5-brane instanton corrections have been taken recently  \cite{Alexandrov:2009vj,APP}. 

Aside from these physics motivations, we hope that our elementary analysis of Eisenstein series for Picard modular groups will be valuable in mathematics, where explicit constructions of automorphic forms are scarce. 

\ack
AK would like to thank the organizers of QTS6 for organizing such a broad conference and the invitation to present these results. The authors are grateful to T.~Finis, J.~Rogawski and J.~Stienstra for correspondence and discussions. AK is a Research Associate of the Fonds National de la Recherche Scientifique-FNRS, Belgium.

\end{document}